# Invisible in Search? Auditing Aesthetic Bias in the Visual Representation of Holocaust Victims on Google

Mykola Makhortykh*, Tobias Rohrbach**, & Maryna Sydorova***

*University of Bern
**University of Fribourg
***University of Bern/University of Fribourg

**Abstract**: Information retrieval systems, such as search engines, increasingly shape the representation of the past and present states of social reality. Despite their importance, these systems face challenges in dealing with the ethical aspects of representation due to various forms of bias, including aesthetic bias that perpetuates hegemonic patterns of representation. While most research on aesthetic bias has examined it in the context of current societal issues, it is also crucial for historical representation, particularly of sensitive subjects such as historical atrocities. To address this gap, we conduct a comparative audit of the visual representation of Holocaust victims on Google. We find that Google tends to propagate a male-dominated representation of Holocaust victims with an emphasis on atrocity context, risking rendering invisible gender-specific suffering and decreasing potential for nurturing empathy. We also observe a variation in representation across geographic locations, suggesting that search algorithms may produce their own aesthetic of victimhood.

**Acknowledgements**: This work was supported by the Alfred Landecker Foundation, which provided funding for the research time of Dr. Makhortykh as part of the project titled "Algorithmic turn in Holocaust memory transmission: Challenges, opportunities, threats", and Cohere, which provided free research credits for the use of its large language models to Dr. Makhortykh as part of its research programme.

**Invisible in Search? Auditing Aesthetic Bias in the Visual Representation of Holocaust Victims on Google**

**Introduction**

Algorithm- and artificial intelligence (AI)-powered systems are ubiquitous elements of today's online platforms, which organise the extensive volumes of information and increasingly shape the representation of present and past states of social reality. The core functionality of these systems, ranging from search engines to chatbots, is the retrieval of information in response to user queries, which affects how individuals perceive and interpret issues for which they seek information (Munton, 2022). However, the degree to which information retrieval systems—including the more advanced ones powered by generative AI (Zhu et al., 2025)—can address the ethical and legal aspects of representation is not entirely clear, especially for highly sensitive subjects (e.g. mass atrocities; Makhortykh, 2023). This uncertainty is particularly concerning when considering the limited reasoning capacities of many information retrieval systems together with their limited ability to understand the context in which representations occur (Merrouni et al., 2019).

One particular example of the challenges faced by information retrieval systems regarding the representation of complex and contested subjects is that these systems are often prone to (algorithmic) bias. Kordzadeh and Ghasemaghaei define bias as a tendency for an algorithmic system to produce outputs that "benefit or disadvantage certain individuals or groups more than others without a justified reason for such unequal impacts" (2022, p. 1). Algorithmic bias can take different forms (Fazelpour & Danks, 2021; Lopez, 2021; Nah et al., 2024), but the one that is particularly relevant in the context of representation is aesthetic bias, which concerns the reiteration of the hegemonic aesthetic patterns (Vianna, 2025), for instance, in the form of stereotypical depictions of individuals and groups (Urman et al., 2022; Vargas-Veleda et al., 2025). Such depictions often exaggerate social expectations regarding the depicted phenomena, reflecting the socially constructed standards of representation (e.g. what is visually appealing or beautiful; Castagna et al., 2021).

Much research on aesthetic bias focuses on the unequal or discriminatory representation of specific population groups in relation to current societal issues (Kay et al., 2015; Noble, 2018; Urman et al., 2022; Rohrbach et al., 2024). However, aesthetic bias can also apply to historical representation, which is essential for a sense of belonging and identity and has strong affective potential. Such a potential results in historical representations often being instrumentalised for public mobilisation, particularly by populist actors, who rely on the verbal and visual aesthetics associated with the past to evoke feelings of nostalgia (Price, 2018) and trauma (Kalstein et al., 2024). In this context, the representation of mass suffering that occurred in the past (e.g. historical genocides) is particularly relevant, both due to its association with strong empathetic reactions and ethical responsibilities to honour the memories of victims.

Considering that information retrieval systems serve as an important source of information about the past (e.g. according to the two surveys conducted by the authors in one of European countries, search engines constitute one of the three most commonly used sources of historical information together with TV shows and online encyclopedias), their aesthetic bias can interfere with the instrumentalisation of the representation of mass atrocities, for instance, by making it easier or harder to associate with the victims depending on how they are portrayed in response to user queries. Importantly, aesthetic bias can also pose challenges for honouring the memory of victims, as it can result in certain groups of victims—e.g. from a specific murder site (Makhortykh et al., 2021)—becoming invisible, thus making them less likely to be remembered.  However, to our knowledge, no study has systematically examined aesthetic bias in the representation of historical mass atrocities and how it varies over time and across different regional contexts.

To address this gap, we look at how the world's most commonly used search engine, Google (Statcounter, n.d.), represents victims of the Holocaust and whether their representation is subject to aesthetic bias. We decided to focus on the Holocaust because it is a particularly well-known instance of historical mass atrocity that spanned the borders of multiple European countries and which shaped representation practices regarding other mass atrocities (Ebbrecht-Hartmann et al., 2023). Because of its significance, memory about the Holocaust and its victims is often instrumentalised both in democratic and authoritarian contexts, with examples ranging from the justification of the military campaigns (Steinweis, 2025) to the public mobilisation for resisting public policies (Ross & Tajima, 2024). Furthermore, considering the multimedia nature of Holocaust memory, understanding how it is depicted by information retrieval systems, particularly search engines that aggregate images from various information sources, is a rather challenging but also urgent task, considering the intensification of appropriation and distortion of Holocaust remembrance following Hamas's attack on Israel on October 7 and Israel's retaliation against Gaza.

The rest of the paper is organised as follows: First, we discuss the aesthetics of Holocaust victimhood and the ways it can influence aesthetic bias in algorithm- and AI-driven information retrieval systems. This is followed by a brief overview of research on aesthetic bias in search engines, which are the focus of the current paper. We then introduce the auditing methodology used to collect search engine data and discuss the analytical approach employed to answer our research questions and test hypotheses. Then, we provide an overview of our findings regarding aesthetic bias in the (1) sociodemographic profile of depicted Holocaust victims, (2) historical context in which victims are portrayed, and (3) geographic context in which searches for Holocaust imagery are conducted.  We then conclude with a discussion of our findings, the limitations of the current study, and directions for future research.

**Aesthetics of Holocaust victimhood**

Like other forms of algorithmic bias, aesthetic bias often emerges from social stereotypes and prejudices, particularly the ones related to societally desirable ways of representing social reality. For representations of historical atrocities, such social desirability can be motivated by various reasons: from facilitating historical knowledge acquisition and understanding (e.g. via the use of visuals for history teaching; Haward, 2020) to making atrocity narratives more appealing for the public, for example, to increase empathy towards past suffering or instrumentalise it for political gains (Sontag, 2003). The process of constructing aesthetic representations of historical atrocities spans diverse forms of media, ranging from traditional ones, such as film (Gelbin, 2011) and literature (Cohen, 1998), to more novel forms, including video games (Ciáurriz, 2023), memes (González-Aguilar & Makhortykh, 2022).

Among many aesthetic aspects of historical atrocity representation, the aesthetics of victimhood are crucial for understanding (but also distorting) memory about past suffering (Sodaro, 2019). Through the representation of victims, the immense and often incomprehensible scale of suffering can be contextualised and made more relatable. Images, such as photographs, drawings, or films, provide evidence of atrocities characterised by "an immediacy and authority greater than any verbal account" (Sontag, 2003, p. 24). It allows future generations to establish a personalised connection with past suffering, moving beyond the statistical representation of an atrocity (e.g. based on the victim count) to visualise individual stories and engage with them. The same features, however, also facilitate instrumentalisation of victimhood in today's political context (Chouliaraki, 2024).

For many historical atrocities, including the Holocaust, it is difficult to identify a single aesthetic canon for victim representation. Despite being a global memory phenomenon (Assmann, 2010), the role and representation of the Holocaust within national and regional memory practices vary broadly. Even in Europe, where the Holocaust took place, there are major differences in the public awareness about the Holocaust and, consequently, varying practices of its public commemoration, both regarding the scope of commemoration and its particular (aesthetic) focus (Dreyfus & Stoetzler, 2011; Hansen-Glucklich, 2016), with even more aesthetic differences in memorial practices found across the globe (Klacsmann, 2024). Furthermore, as Holocaust remembrance is a dynamic process, the representation of victims evolves over time and can be influenced by various factors, including changes in memory regimes, the discovery of new historical evidence, and the emergence of new technologies for representing the past. Lastly, an additional challenge of the Holocaust case is that it is represented via an extremely diverse range of media, making it difficult to identify a single aesthetic principle that persists across these diverse spaces.

These difficulties explain why, to our knowledge, there is no comprehensive assessment of the aesthetics of Holocaust victimhood across different media and different countries. However, based on existing research, we can still identify certain commonalities that can shape aesthetic bias in web search engines. It is important to note that these observations are particularly applicable to the aesthetic aspects of memorial practices associated with Holocaust heritage institutions in Western Europe and the United States, whereas in other regions they may differ significantly (Petö & Klacsmann, 2025). However, these practices are also the ones that are likely to have a stronger impact on the algorithm- and AI-based representation of Holocaust victims, as they are promoted by more well-known institutions, which have greater resources to enhance their visibility, as evidenced by earlier research on algorithmic representation of the Holocaust (Makhortykh et al., 2023).

The first commonality relates to the selective sociodemographic construction of victimhood, characterised by a strong focus on *feminine representation of victimhood* based on Western standards of beauty (Kaplan, 2019; Głowacka, 2020), together with emphasis on more vulnerable and sympathetic groups of victims, particularly children and women (Jacobs, 2008). Analysing Holocaust liberation photographs, Zelizer (2001, p. 251) states that "images of women filled the pages of the daily and weekly press" and that "evidence of gender literally leaked from the photographs", even while the degree to which such representation translated into the adequate representation of female experiences of the Holocaust remains questioned (Parks, 2025).

There is also a growing emphasis within Holocaust memorial practices on the third generation of victims, namely the grandchildren of survivors (e.g. Apostolo, 2025), who are often shown as youngsters and young adults. In addition to stressing the resilience of survivor families (in the case of the third generation representations), such emphasis on vulnerable groups has the strategic purpose of contrasting with images of (male) brutality. In the context of Google image search, such patterns can translate into certain groups of victims, for instance, middle-aged men, being less likely to appear as victims, whereas women and children can become disproportionately overrepresented. We hence formulate our first hypothesis:

**H1**: Search results will contain more images of (a) adult women and (b) children than of adult men

The gendered nature of victimhood is likely to also manifest in gender asymmetries in emotional expressions. Research using the Stereotype Content Model (Fiske et al., 2002) suggests that women, especially in contexts of suffering or caregiving, tend to be stereotyped as high in warmth but lower in competence, reinforcing traditional gender roles (Cuddy et al., 2008). In Holocaust imagery, this dynamic may be amplified by the gendered framing of victimhood, where female victims are more readily associated with emotional vulnerability and passive suffering (Chouliaraki, 2021; Zelizer, 2001). Together with the known tendency of search algorithms to reiterate emotional stereotypes in gender representation (Otterbacher et al., 2017), Googly may disproportionately depict female Holocaust victims with warmth-related expressions (e.g. sadness, empathy), while reserving competence-related emotions (e.g. resilience, defiance) for men. We therefore expect:

**H2**: Search results will be (a) more likely to depict women with warmth-related emotions than men and (b) less likely to depict women with competence-related emotions than men

Another commonality of victimhood representation that we expect to be relevant for the algorithmic gaze of Google is a shifting focus on specific historical environments in which victims are represented. While much attention after the end of the Second World War has been paid to the concentration and extermination camps and the presence of victims there (either as of liberated living individuals or as traces of those who perished), in recent decades, there has been a tendency to move away from the explicit focus on *gruesome depictions* of the victims through the prism of atrocities, especially since many of these depictions were made against victims' will and come from the perpetrators. Rather than concentrating exclusively on the experiences of violence and death (and confining victims of the Holocaust to a very particular and narrow role as illustrations of human suffering), the representation of victims also considers other, less explicit aspects of the Holocaust. Simultaneously, both in museums and popular culture, there is a stronger emphasis on the time preceding the Holocaust, showcasing the pre-war life of victims, and following the Holocaust, with a focus on the post-war life of survivors and their descendants. It leads us to the following expectation regarding a potential shift towards a more nuanced Holocaust representation:

**H3**: Search results will include fewer images focusing on victims in the context of the atrocity rather than outside of it

**Aesthetic bias in search engine outputs across the world**

Search engines are key information intermediaries in today's media environment, and have a significant impact on the representation of social reality. Increasingly, search engines, such as Google, use AI to interpret user queries and select the most relevant information sources (Reid, 2025); by doing so, they determine what interpretations of societally relevant phenomena individuals are exposed to. These interpretations are highly trusted by individuals who typically perceive search engines as reliable sources of information (Schultheiß & Lewandowski, 2023) and extensively use them (Urman & Makhortykh, 2023), even while advances in generative AI, in particular the emergence of AI-powered chatbots capable of retrieving information online, will likely pose a major challenge for search engines' business models.

The importance of search engines underscores the need to comprehend how their representation of social reality can be susceptible to various forms of algorithmic bias, including aesthetic bias. Multiple studies have examined the gendered aspects of aesthetic bias, particularly in the representation of vulnerable groups, such as women (Kay et al., 2015; Otterbacher et al., 2017), people of colour (Noble, 2018), and population groups outside the Global North (Urman & Makhortykh, 2024). The specific areas in which the presence and effects of such bias were studied varied from technological innovation (Makhortykh et al., 2021) to professional occupations (Kay et al., 2015) to politics (Pradel, 2021; Rohrbach et al., 2024; 2025).

Despite research on aesthetic bias on search engines steadily growing, studies on the potential skewness in aesthetic representation of historical atrocities by search engines remain very few. A few exceptions (Makhortykh et al., 2021, 2024) examined the case of the Holocaust and observed a tendency for search engine outputs to be dominated by a few prevalent visual themes, such as images of the liberated camps, often sourced from a few well-known Holocaust-related sites. There were also indicators that aesthetic bias may vary depending on the language of the query and change over time, for instance, with search engines increasingly prioritising images of modern memory sites instead of historical photos (Makhortykh et al., 2024). However, to our knowledge, no study has so far looked at how search engines aesthetically represent victims of either the Holocaust or any other genocide.

One of the reasons why our understanding of search-based aesthetic bias regarding historical atrocities remains extremely limited is due to the multitude of factors which affect search outputs. In addition to the time-based changes in representation of specific phenomena, which are due to the evolution in source relevance (Hannak et al., 2013; Makhortykh et al., 2024), search engines are influenced by how queries are formulated. However, even the same queries in different languages can be quite different, as search engines seek relevant results within the pool of pages written in that respective language (Google, 2023). This search localisation is further coupled with personalisation based on the user's location, aiming to promote geographically relevant search results (Kliman-Silver et al., 2015).

Under these circumstances, we expect that the language of the query and the location from which the search is conducted can be key factors for the emergence of aesthetic bias. This expectation is based on earlier studies highlighting the significant role of search localisation on aesthetic representation of specific phenomena, particularly regarding gendered representation (Vlasceanu & Amodio, 2022; Rohrbach et al., 2024). However, it remains unclear which factors in the case of representation of victimhood in the context of mass atrocities can explain the geographical variation. One of them can be related to the differences in representation of the Holocaust in a specific national context that we have noted earlier; its immediate consequence is that for the specific locations and specific language pools, search engines will have to operate with the different sets of images from which they will select more relevant results. While it is difficult to identify detailed expectations based on the rather different practices of Holocaust representation, we suggest that the number of Holocaust victims originating from specific countries and the number of Holocaust memorial sites can be indicative of a higher relevance of Holocaust memory. Such higher relevance is usually accompanied by the active integration in global Holocaust memory practices, characterised by the tendency to humanise genocide remembrance (both in terms of giving more visibility to victims as human beings and not abstract symbols and not limiting their representation to their immediate atrocity-related experience). Hence, we formulate the following expectations:

**H4**: The number of images with humans will be higher for countries (a) with more Holocaust memorial sites and (b) from which more Holocaust victims originated

**H5**: The number of images showing victims in the context of atrocities will be lower for countries with more Holocaust memorial sites

Another factor which we consider relevant for explaining differences in aesthetic bias depending on the search location is gender equality. The state of gender equality has been shown to be highly relevant to many aspects of gendered outcomes (Stoet & Geary, 2018), including the representation of specific gender groups in both the present and the past (Vlasceanu & Amodio, 2022; Rohrbach et al., 2024). Despite the highly gendered nature of the aesthetics of Holocaust victimhood, we expect that the gender equality index can influence the variation in the visibility of men and women in the victimhood-related search results, leading us to the following expectation:

**H6**: The present-day gender equality score will influence the distribution of female and male victims depicted in image search results.

**Methodology**

*Data collection*

To test our hypotheses, we conducted an algorithm audit of the visual representation of Holocaust victims in Google image search. As noted earlier, we focused on Google because it is the most commonly used search engine in most parts of the world. To conduct an algorithm audit, that is, a systematic examination of "functionality and/or impact of algorithmic systems" (Urman et al., 2025, p. 375), we manually collected the top 50 search results from 35 locations worldwide for a gender-neutral queries - "Holocaust victim" and "Holocaust victims" - into the official languages of the country (and, in some cases, the second most commonly used language) corresponding to each location. The majority of chosen locations were in Europe, where the Holocaust took place (for the complete list of locations and corresponding queries, see Appendix 1). We simulated the location using the virtual private service provided by Le VPN; before each search, we tested if Le VPN provided an internet protocol address corresponding to the declared location and included only those locations where the match happened. To prevent the possible effects of search personalisation, we used a clean browser for data collection, which was cleaned after each search query. Overall, we collected 4,381 image search results (the number is due to the variation in the number of images included in search pages for a specific language-location combination).

*Data processing*

To enrich image search results, we used computer vision to automatically label variables of interest. To identify the gender (treated as binary male vs. female; see the discussion for more reflection on this limitation), the estimated age (in years), and emotional expressions of individuals depicted in image search results, we used the Amazon Rekognition application programming interface, a state-of-the-art computer vision software with extensive capabilities for facial expression analysis. Despite the risk of computer vision software being prone to bias, earlier studies (e.g. Ulloa et al., 2024) have demonstrated that the performance of Amazon Rekognition is comparable to human assessments, specifically in identifying human-related features (e.g. the number of individuals and their gender). To identify the context in which individuals are shown (i.e. the context of atrocity vs. non-atrocity context), we utilised the c4ai-aya-vision-32b model developed by Cohere. We used this state-of-the-art multimodal model to determine if images show people who are likely to be imprisoned, escorted to be deported or arrested, incarcerated in a prison camp, or dead (which was our definition of the atrocity context). To validate the performance, we tested the model's performance on a golden standard manually labelled by one of the authors, who is an expert in Holocaust memory research; the model achieved a rather high result of 0.89 macro F1-score.

To operationalise the above-mentioned characteristics for analysis, we employed the following approach. From Amazon Rekognition results, we extracted the predicted *gender* (male or female) for all individuals in specific images as well as their *estimated age* (in years). We measured the *presence of minors* by recoding the estimated age into a dummy variable indicating whether the depicted person is of legal age (0 = 18 years or older; 1 = 18 years old or younger). We also identified nine emotional expressions (see below) of depicted persons as probabilistic prevalence scores from 0 (emotional expression is certain to be absent) to 1 (emotional expression is certain to be present). We then compared the scores for each emotion for each image to make prevalence estimates of the emotions' presence for each image. We used the SCM (Fiske et al., 2002) to group the facial expressions into *warmth-related* (happy, smile, calm), *competence-related* (angry, disgusted), and *other emotions* (sad, confused, surprised, fear).

Finally, we coded several characteristics at the query level. We marked the query's *country*, namely, from which of 35 countries it was used (see Table A1 in the Appendix for an overview). For each country, we aggregated image data to measure both the *share of images with at least one human* and the *average share of women* in image search results. Finally, we enriched the dataset with contextual characteristics of the countries by adding three extramedia variables from different data sources to test hypotheses regarding the effects of search localisation on aesthetic bias (H4-H6). First, we used the International Holocaust Remembrance Alliance database (IHRA, n.d.) to identify the *number of Holocaust memorial sites* in each country. Second, we added the *number of the individual Holocaust victims* per country (specifically, the number of those who perished) based on the estimates provided by the United States Holocaust Memorial Museum (USHMM, n.d.). Third, we operationalised countries' overall *gender equality* using the Global Gender Gap Index (GGGI) from the 2024 report by the World Economic Forum (WEC, 2024). A GGGI score of 1 indicates gender equality (i.e. no gender gap).

**Findings**

*Patterns of aesthetic bias in Holocaust search results*

We first tested for aesthetic bias in Google searches of the Holocaust across sociodemographic variables. Contrary to our expectation (H1a), a multilevel logistic regression model with country-level random intercepts indicated that adult women are less visible in search results than men ($M$ = 0.403, $SE$ = 0.007, $p$ < .001; 95% CI = [0.394, 0.424]; see model 1 in Table 1). Panel A of Figure 1 shows women's average presence in search results by country (blue bars), indicating minimal cross-national variation. Notably, women's representation never reaches or exceeds 50%, providing consistent evidence against our expectations. The same panel also reveals the share of depicted children (i.e. individuals under 18 years old; purple bars), with an average presence of 14.4% (95% CI = [0.131, 0.158]). When employing a linear operationalisation of depicted persons' age rather than a dummy variable, the mean estimated age was found to be 28.4 years (95% CI = [27.9, 29.0]; see model 2 in Table 1). These results provide no support for our expectation that search results overrepresent children (H1b), although Google still depicts Holocaust victims as relatively young.

In light of the findings contradicting our hypotheses, we examined whether the aesthetic focus on women and children—while not applicable to Google's representation of Holocaust victims in general—might manifest as an interaction between age and gender. Panels B and C of Figure 1 reveal an interesting pattern for two different analyses. Panel B shows the predicted marginal means for a significant gender effect on the presence of minors ($b$ = -0.795, $SE$ = 0.031, $p$ < .001): 20.1% of minors are girls (95% CI = [0.183, 0.217]), compared to approximately 10.1% who are boys (95% CI = [0.092, 0.112]). Panel C offers another perspective by illustrating the effect of the estimated age of depicted individuals on the share of women in images ($b$ = -0.024, $SE$ = 0.001, $p$ < .001). The relatively strong negative association suggests that women are overrepresented in depictions of young individuals (under 23 years old) but are underrepresented among adults. This finding nuances the null findings for H1, highlighting that while Google tends to feature predominantly older men for queries about Holocaust victims, women are prevalent among younger victims, thus reflecting a complex representation of vulnerability and victimhood.

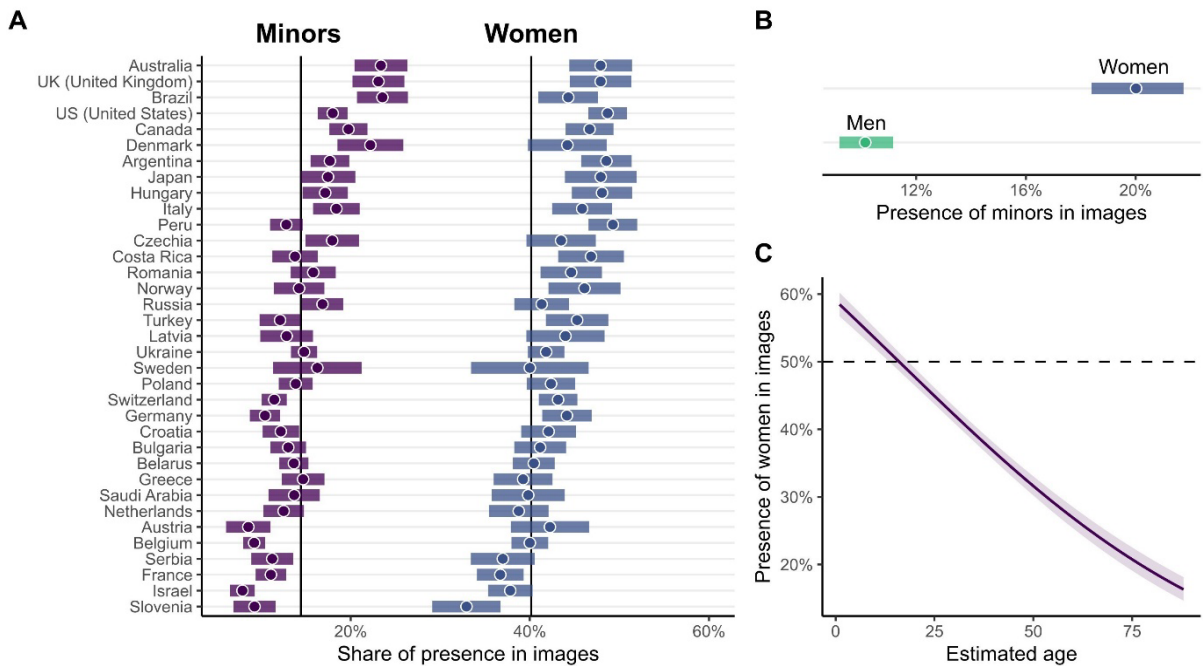

**Figure 1.** Overview of the prevalence of minors and adult women in Holocaust-related Google searches. Panel A displays the descriptive prevalence by country of query, separated for women and minors, while Panels B and C illustrate the interaction between the two social categories.

To test H2 regarding gender differences in emotional expressions, we conducted independent samples t-tests comparing women and men across nine facial emotions, categorised into warmth-related (happy, smile, calm), competence-related (angry, disgusted), and other emotions (sad, confused, surprised, fear). To control for multiple comparisons, we applied the Bonferroni correction across all nine tests. In line with hypothesis H2a, women consistently displayed more warmth-related emotions than men. Specifically, women in search results were more likely to show happiness expressions ($M = 12.5$, $SD = 27.4$) compared to men ($M = 8.4$, $SD = 21.4$), $t(28662) = 15.2$, $p < .001$, d = 0.16. Women were also more likely to be shown as smiling ($M = 10.5$, $SD = 30.7$) than men ($M = 6.9$, $SD = 25.3$), $t(29742) = 12.0$, $p < .001$, $d = 0.13$. Against our expectation, however, women showed significantly lower calm expressions ($M = 41.8$, $SD = 34.7$) compared to men ($M = 48.3$, $SD = 35.7$), $t(33947) = -17.1$, $p < .001$, $d = -0.18$. Note that the magnitude of these gender differences is overall small.

Hypothesis H2b predicted women would display fewer competence-related emotions than men. This hypothesis received no support after correcting for multiple comparisons. Descriptively, women had more expressions of anger ($M = 5.1$, $SD = 14.5$) and disgust ($M = 1.7$, $SD = 4.4$) than men (anger: $M = 4.8$, $SD = 13.6$; disgust: $M = 1.5$, $SD = 4.9$), but these differences were not statistically significant, $t(32369) = 1.94$, $p = .472$, $d = 0.02$ and $t(35070) = 2.43$, $p = .136$, $d = 0.03$. The two largest differences between men and women emerged in other emotional categories, with women showing more fearful ($d = 0.17$) and less sad expressions ($d = 0.26$) compared to men (all $p < .001$). In turn, search engines yielded more images with men displaying more confusion than women ($d = -0.09$, $p < .001$), while surprise showed no significant gender difference ($p = .98$).

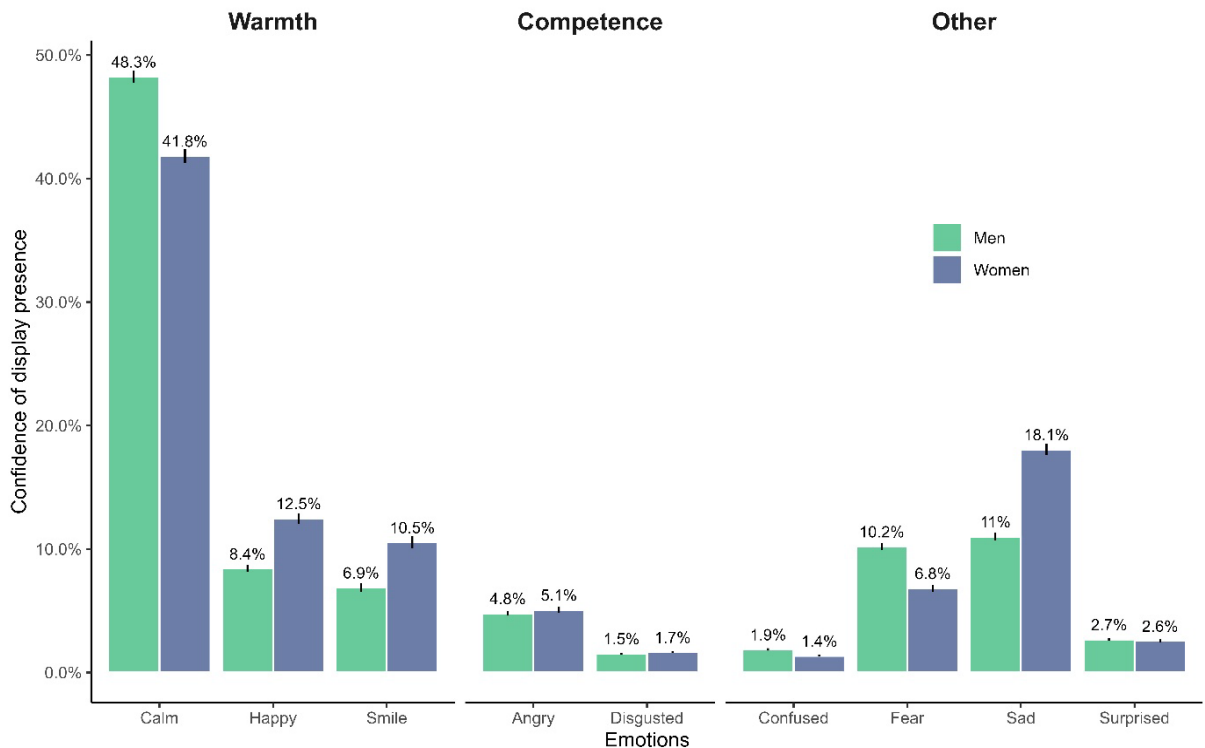

**Figure 2.** Visualisation of the average prevalence of warmth-related, competence-related, and other emotional displays for women and men depicted in Holocaust-related queries.

We then tested aesthetic bias in the visibility of atrocities in relation to the depiction of Holocaust victims. To test hypothesis H3 that search results would include fewer images showing victims in the context of atrocity, we conducted a multilevel logistic regression with country as a random effect (see model 3 in Table 1). The analysis suggested that atrocity-related content was present in 75.7% (95% CI [69.5%, 81.1%]) of results across all countries. An exploratory analysis revealed a significant interaction between gender and age regarding the presence of victims in the atrocity context ($b$ = -0.087, $SE$ = 0.027, $p$ = .001). Women showed a steeper negative slope ($b$ = -0.520), indicating that the presence of atrocity decreased more rapidly for them with age compared to men ($b$ = -0.608). This suggests that while both genders show declining representation in atrocity-related images with increasing age, the gap in atrocity presence narrows as age increases. The interaction demonstrates that younger women are more likely to be shown in the context of atrocity than younger men, but this difference diminishes among older individuals.

**Table 1**. Full regression results of main models predicting the gender and estimated age of depicted persons as well as atrocity-related context in Google searches of the Holocaust

| | Model 1: Presence women | | | Model 2: Estimated age | | | Model 3 Presence atrocity | | |
|---|---|---|---|---|---|---|---|---|---|
| *Predictors* | *OR* | *SE* | *p* | *B* | *SE* | *p* | *OR* | *SE* | *p* |
| (Intercept) | 2.83 | 1.38 | **0.033** | 26.57 | 4.54 | **<0.001** | 3.02 | 0.48 | **<0.001** |

*Image-level predictors*

| | Est | SE | p | Est | SE | p | Est | SE | p |
|---|---|---|---|---|---|---|---|---|---|
| Confidence Gender classification | 0.22 | 0.02 | **<0.001** | | | | | | |
| Gender [man] | | | | 3.48 | 0.13 | **<0.001** | 1.19 | 0.03 | **<0.001** |
| Estimated age | 0.98 | 0.00 | **<0.001** | | | | 0.59 | 0.01 | **<0.001** |
| Gender [man] × Age | | | | | | | 0.92 | 0.02 | **0.001** |
| Presence of atrocity | 0.86 | 0.02 | **<0.001** | | | | | | |

*Query-level predictors*

| | Est | SE | p | Est | SE | p | Est | SE | p |
|---|---|---|---|---|---|---|---|---|---|
| GGGI Score | 2.04 | 1.35 | 0.280 | 0.94 | 6.22 | 0.879 | 0.97 | 0.19 | 0.870 |
| Memorial sites | 0.99 | 0.03 | 0.725 | -0.26 | 0.27 | 0.335 | 0.91 | 0.13 | 0.545 |
| Number of victims (log) | 0.99 | 0.01 | **0.040** | -0.01 | 0.06 | 0.925 | 1.08 | 0.20 | 0.670 |

**Random Effects**

| | | | |
|---|---|---|---|
| $\sigma^2$ | 3.29 | 140.16 | 3.29 |
| $\tau_{00}$ | 0.03 Country | 2.79 Country | 0.87 Country |
| ICC | 0.01 | 0.02 | 0.21 |
| N | 35 Country | 35 Country | 35 Country |
| Observations | 30789 | 36009 | 35997 |
| Marg. /Cond. $R^2$ | 0.031 / 0.039 | 0.021 / 0.040 | 0.075 / 0.268 |

*Cross-national variation in Holocaust search results*

The remaining expectations investigated the relationship between countries' contextual characteristics of Holocaust remembrance and the visual depiction of the Holocaust on Google searches. We expected that the number of images with humans in search results would be higher for countries with more Holocaust memorial sites (H4a) and for countries from which more Holocaust victims originated (H4b). Correlation analyses at the country level (N = 35 countries) provided no support for this expectation. Specifically, we found no significant association between the number of Holocaust memorial sites in a country and the share of images containing humans in Google search results, $r$ = -.07, $n$ = 35, $p$ = .679 (Panel A of Figure 3). Moreover, we observed a moderate *negative* correlation between the number of Holocaust victims from a country and the share of images with humans in search results, $r$ = -.41, $n$ = 35, $p$ = .015 (Panel B of Figure 3). This indicates the exact opposite relationship than hypothesised: Queries in countries with higher historical victim counts were less likely to produce search results featuring human subjects, accounting for roughly 17% of the variance in human representation.

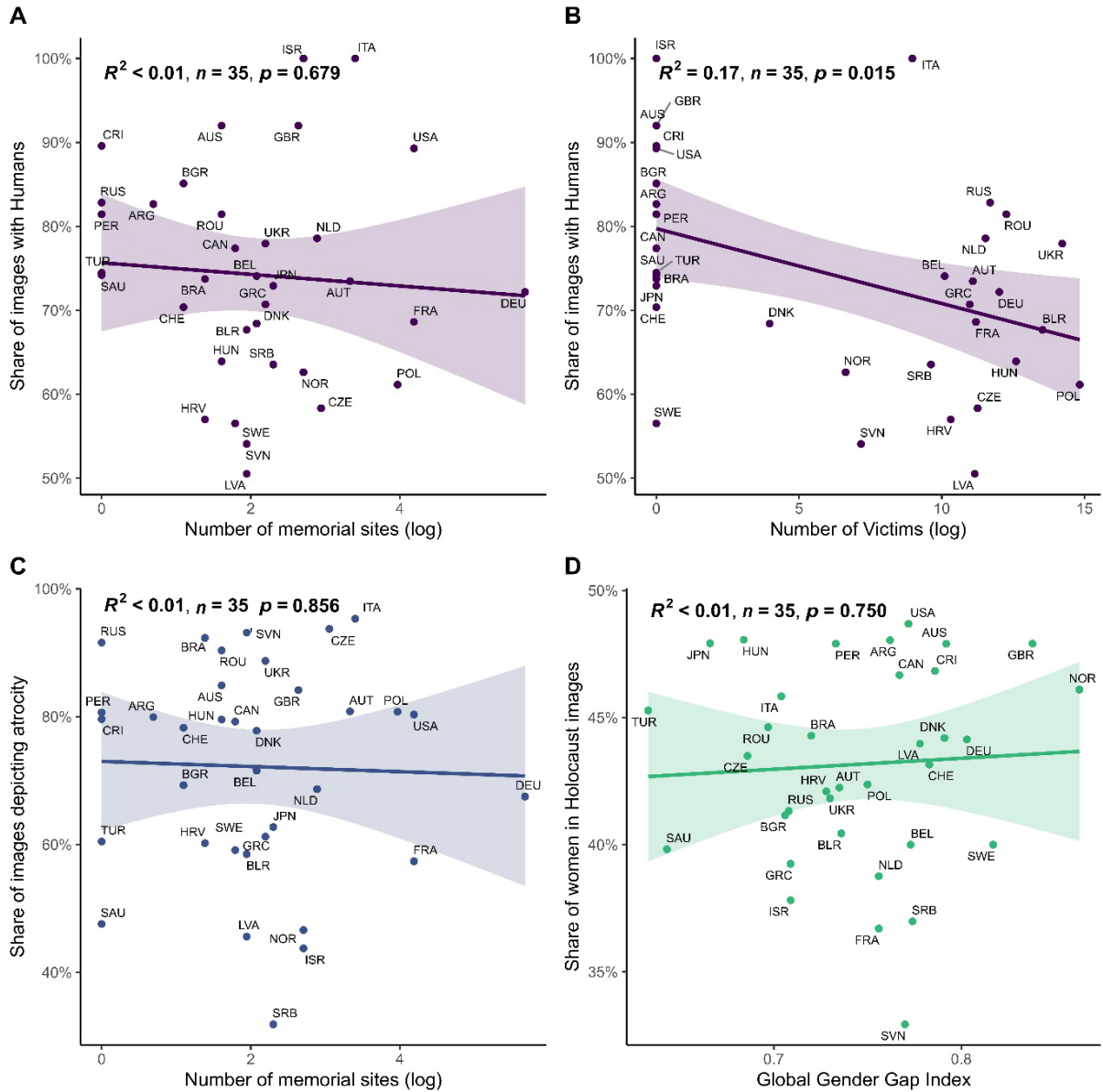

**Figure 3**. Illustration of bivariate correlations between image-level (y-axes) and query-level (x-axes) characteristics of Holocaust image searches

The remaining expectations regarding cross-national variation in visual Holocaust Google queries did not receive empirical support. Specifically, we did not observe the anticipated negative association between the share of atrocity-related images and the number of Holocaust memorial sites in a country (H5), $r < -0.03$, $n = 35$, $p = .856$ (see Panel C of Figure 3). Additionally, our final expectation (H6) was also unsupported, as the results indicated no significant association between the countries' contemporary global gender gap index and the share of women depicted in Holocaust image queries, $r < 0.06$, $n = 35$, $p = .679$ (see Panel D of Figure 3).

**Discussion and Conclusion**

In this article, we examined how Google, the world's most used search engine, visually represents Holocaust victims and whether such representation is subject to aesthetic bias. Through a large-scale algorithm audit of Google's image search, we explored results for the gender-neutral queries "Holocaust victim" and "Holocaust victims" in different languages from 35 locations, including both countries where the Holocaust occurred and countries which were not directly affected by it. Using computer vision techniques, we enriched the collected images and tested a selection of hypotheses informed by the research on the aesthetics of Holocaust representation and memory, as well as search engine bias.

Our key finding is that, except for the tendency to associate women with warm emotions—that is, a common form of aesthetic bias in algorithmic gender representation in other domains (Otterbacher et al., 2017; Rohrbach et al., 2025)—none of our hypotheses regarding aesthetic bias has found support. We did not find evidence of Google reiterating the aesthetic bias associated with established patterns of victimhood in general, or Holocaust victimhood in particular. Our expectations regarding the relationship between the representation of victims and the gender equality index, or the significance of Holocaust memorial culture in specific regions, were also not supported.

However, the lack of support for specific forms of aesthetic bias, which we expected to find in Google search results, does not imply that the algorithmic representation of Holocaust victims is not skewed towards certain aesthetic patterns. In the algorithmic gaze of Google, Holocaust victims appear as primarily older males, usually shown in the context of atrocity (e.g. being deported or in the concentration camp setting). It aligns with the earlier observations regarding the prevalence of male-oriented representation for gender-neutral queries on search engines (Urman et al., 2022; Vlasceanu & Amodio, 2022). Most images also featured humans (as contrasted, for instance, by showing only memorial sites without a human presence), thus giving a human—and predominantly male—face to Holocaust victimhood.

Importantly, there was rather substantive variation in victims' representation for specific combinations of queries and locations. For instance, in response to queries in Hebrew from Israel, all images featured humans, whereas for Sweden or Croatia, the proportion of images offering a human face to the victimhood was less than 60%. Similarly pronounced was the variation in the presence of images showing victims in the immediate context of atrocity (e.g. of victims being in the concentration camps): for Italian queries from Italy, it was close to 100%, whereas for Serbian and Hebrew queries, less than 50% of queries featured the atrocity context. Such broad variation highlights the importance of search localisation (Kliman-Silver et al., 2015) for representing the present and past states of social reality, as well as the degree to which such representation can be skewed (and the directionality of this skewness). While it is difficult to identify a particular pattern, it seems that countries not affected by the Holocaust were more likely to include images with humans and also with women

These patterns and their divergence from our expectations can be attributed to several factors. First, they may be due to the difference between various aspects of Holocaust memorial culture: while in physical memorials and historical reporting, female-focused aesthetics has been prevalent (Jacobs, 2008; Zelizer, 2001), the representation of the Holocaust in some memorial institutions has been criticised for the limited visibility of women (Parks, 2025; Petö & Klacsmann, 2025). While our expectation was that Google's representation of victimhood would be affected primarily by the former tradition, it may be that it is more reliant on aesthetic patterns associated with more male-oriented institutional practices. Second, Google search results are shaped not only by memory practices in the specific region and their representation by heritage institutions, but also by other sources (e.g. journalistic media), which may not necessarily follow the same aesthetic principles of victimhood representation. Third, the images retrieved by Google may include not only victims, but also potentially perpetrators or other individuals (e.g., present-day politicians attending commemorative events). While the presence of these individuals would still shape user perceptions of the aesthetics of victimhood, it may cause misalignment with our expectations based on aesthetics emerging in a more carefully human-curated environment.

Together, however, our findings suggest that increasingly AI-driven information curation on Google produces its own aesthetics of Holocaust victimhood. Such aesthetics is rather different from a traditional focus on women and children, which have been historically viewed as emblematic victims (Jacobs, 2008) and whose visibility (in the case of women; Petö & Klacsmann, 2025) has been growing in recent representation practices related to Holocaust memory. Instead, Google tends to show Holocaust victims as older males, usually in the context of atrocity, thus obscuring more gender-specific crimes (and related forms of female suffering) and making it harder to express empathy (which is one of the reasons modern memorial practices portray victims beyond the context of atrocity). Importantly, this aesthetic also varies strongly with the location and the language of the user query, which, considering the importance of search engines for information-seeking, may result in a rather different perception of both the concept of victimhood and the atrocity to which it relates.

Finally, it is important to note several limitations of the current study. The first of them is that we did not conduct additional verification of images to identify which of them actually feature Holocaust victims and which show other types of content (e.g. modern politicians delivering speeches at the Holocaust sites). Second, our operationalisation of the factors which can influence the geographic aspects of bias is rather simplistic and future research can consider more nuanced ways of classifying different Holocaust memory cultures and evaluating whether they match specific patterns of algorithmic representation. Third, although other research suggests that the use of computer vision for automated gender detection in images is reliable (Rohrbach et al., 2024), this approach is restricted to a binary classification, which overlooks the intricacies of gender identity (Scheuerman et al., 2019).

Finally, this audit concentrates solely on the representation of Holocaust victimhood by Google. Though widely used, it is not necessarily representative of other search engines or AI-driven platforms. Future studies should expand their scope to include other platforms, such as TikTok, which actively utilise AI-powered information retrieval to organise visual content, particularly as there is evidence that it leads to the promotion of harmful political messages (Weimann & Masri, 2023). Similarly, future research can benefit from looking at new types of information retrieval and generation systems, such as digital duplicates—the AI-enabled copies of real humans—which show substantial promise for transmitting memories of mass atrocities (Kozlovski & Makhortykh, 2025). Despite these limitations, this study underscores the importance of critically examining AI-driven aesthetics in the representation of sensitive historical events, as they not only shape public perceptions of victimhood but also play a crucial role in influencing collective memory in an increasingly digital world.

# References

Apostolo, S., Kohlbauer-Fritz, G. & Meisinger, A. (2024). *Die Dritte Generation. Der Holocaust im familiären Gedächtnis*. Hentrich & Hentrich Verlag.

Assmann, A. (2010). The Holocaust—A global memory? Extensions and limits of a new memory community. In A. Assmann & S. Conrad (Eds.), *Memory in a global age: Discourses, practices and trajectories* (pp. 97-117). Palgrave.

Castagna, A., Pinto, D., Mattila, A., & de Barcellos, M. (2021). Beauty-is-good, ugly-is-risky: Food aesthetics bias and construal level. *Journal of Business Research, 135*, 633-643.

Cohen, R. (1998). The political aesthetics of Holocaust literature: Peter Weiss's The Investigation and its critics. *History & Memory, 10*(2), 43-67.

Chouliaraki, L. (2021). Victimhood: The affective politics of vulnerability. *European Journal of Cultural Studies, 24*(1), 10-27.

Chouliaraki, L. (2024). *Wronged: The weaponization of victimhood*. Columbia University Press.

Ciáurriz, I. (2023). Playing with the unspeakable: The Holocaust and videogames. In M. Mandal & P. Das (Eds.), *Holocaust vs. popular culture* (pp. 48-59). Routledge.

Cuddy, A. J., Fiske, S., & Glick, P. (2008). Warmth and competence as universal dimensions of social perception: The stereotype content model and the BIAS map. *Advances in Experimental Social Psychology, 40*, 61-149.

Dreyfus, J.-M., & Stoetzler, M. (2011). Holocaust memory in the twenty-first century: between national reshaping and globalisation. *European Review of History—Revue européenne d'histoire, 18*(01), 69-78.

Ebbrecht-Hartmann, T., Stiassny, N., & Henig, L. (2023). Digital visual history: Historiographic curation using digital technologies. *Rethinking History*, 27(2), 159-186.

Fazelpour, S., & Danks, D. (2021). Algorithmic bias: Senses, sources, solutions. *Philosophy Compass, 16*(8), 1-16.

Fiske, S. T., Cuddy, A. J., Glick, P., & Xu, J. (2002). A model of (often mixed) stereotype content: Competence and warmth respectively follow from perceived status and competition. *Journal of Personality and Social Psychology*, 82(6), 878-902.

Gelbin, C. (2011). Cinematic representations of the Holocaust. In J.-M. Dreyfuss & D. Langton (Eds.), *Writing the Holocaust* (pp. 26-41). Bloomsbury.

Głowacka, D. (2020). Gendered representations of beauty and the female body in Holocaust art. *Miejsce, 6*, 1-21.

González-Aguilar, J., & Makhortykh, M. (2022). Laughing to forget or to remember? Anne Frank memes and mediatization of Holocaust memory. *Media, Culture & Society, 44*(7), 1307-1329.

Google. (2023, September 8). *How Google Search handles multilingual searches*. https://developers.google.com/search/blog/2023/09/multilingual-searches

Hannak, A., Sapiezynski, P., Molavi Kakhki, A., Krishnamurthy, B., Lazer, D., Mislove, A., & Wilson, C. (2013). Measuring personalization of web search. In *Proceedings of the 22nd International Conference on World Wide Web* (pp. 527-538). ACM.

Hansen-Glucklich, J. (2016). Poetics of memory: Aesthetics and experience of Holocaust remembrance in museums. Dapim: Studies on the Holocaust, 30(3), 315-334.

Haward, T. (2020). How do students engage with visual sources in the teaching and learning of History? *British Educational Research Journal*, *46*(2), 364-378.

IHRA. (n.d.). *Memorial museums database*. Retrieved November 25, 2025, from https://holocaustremembrance.com/resources/memorial-museums-database

Jacobs, J. (2008). Gender and collective memory: Women and representation at Auschwitz. *Memory Studies*, *1*(2), 211-225.

Kalstein, J. et al. (2024). Weaponising collective trauma: The case of Russia and Israel. *Cambridge Journal of Political Affairs, 5*(2), 79-98.

Kaplan, M. (2019). Did gender matter during the Holocaust? *Jewish Social Studies*, 24(2), 37-56.

Kay, M., Matuszek, C., & Munson, S. (2015). Unequal representation and gender stereotypes in image search results for occupations. In *Proceedings of the 33rd annual ACM conference on human factors in computing systems* (pp. 3819-3828). ACM.

Klacsmann, B. (2024). Invisibilizing responsibility: The Holocaust museums of Slovakia and Hungary. *Eastern European Holocaust Studies*, 2, 303-328.

Kliman-Silver, C., Hannak, A., Lazer, D., Wilson, C., & Mislove, A. (2015). Location, location, location: The impact of geolocation on web search personalization. In *Proceedings of the 2015 Internet Measurement Conference* (pp. 121-127). ACM.

Kordzadeh, N., & Ghasemaghaei, M. (2022). Algorithmic bias: Review, synthesis, and future research directions. *European Journal of Information Systems*, *31*(3), 388-409.

Kozlovski, A., & Makhortykh, M. (2025). Digital Dybbuks and virtual Golems: the ethics of digital duplicates in Holocaust testimony. *Memory, Mind & Media, 4*, 1-20.

Lopez, P. (2021). Bias does not equal bias: A socio-technical typology of bias in data-based algorithmic systems. *Internet Policy Review, 10*(4), 1-29.

Makhortykh, M. (2023). No AI after Auschwitz? Bridging AI and memory ethics in the context of information retrieval of genocide-related information. In A. Chakraborty, S. Kumar, A. Mukherjee, & J. Kulshrestha (Eds.), *Ethics in artificial intelligence: Bias, fairness and beyond* (pp. 71-85). Springer.

Makhortykh, M., Urman, A., & Ulloa, R. (2021). Hey, Google, is it what the Holocaust looked like? Auditing algorithmic curation of visual historical content on Web search engines. *First Monday*, *26*(10), 1-24.

Makhortykh, M., Urman, A. and Ulloa, R. (2021). Detecting race and gender bias in visual representation of AI on web search engines. In *International Workshop on Algorithmic Bias in Search and Recommendation* (pp. 36-50). Springer.

Makhortykh, M., Urman, A., Ulloa, R., Kulshrestha, J. (2023). Can an algorithm remember the Holocaust? Comparative algorithmic audit of Holocaust-related information on search engines. In I. Groschek & H. Knoch (Eds.), *Digital Memory: Neue Perspektiven für die Erinnerungsarbeit* (pp. 79-93). Wallstein Verlag.

Makhortykh, M., Urman, A., Ulloa, R., Sydorova, M., & Kulshrestha, J. (2024). Does it get better with time? Web search consistency and relevance in the visual representation of the Holocaust. In E. Pfanzelter, D. Rupnow, É. Kovács & M. Windsperger (Eds.), *Connected histories: Memories and narratives of the Holocaust in digital space* (pp. 13-33). De Gruyter.

Merrouni, Z., Frikh, B., & Ouhbi, B. (2019). Toward contextual information retrieval: A review and trends. *Procedia Computer Science, 148*, 191-200.

Munton, J. (2022). Answering machines: How to (epistemically) evaluate a search engine. *Inquiry*, 1-29.

Nah, S., Luo, J., & Joo, J. (2024). Mapping scholarship on algorithmic bias: Conceptualization, empirical results, and ethical concerns. *International Journal of Communication, 18*, 548–569.

Noble, S. (2018). *Algorithms of oppression.* New York University Press.

Parks, K. (2025, March 28). Women's Representation in Holocaust Museums https://holocaustcenter.org/womens-representation-in-holocaust-museums-a-research-brief-by-dr-katie-chaka-parks/

Petö, A., & Klacsmann, B. (2025). From unremembered to overremembered. Gender in the Holocaust museums of Hungary and Slovakia. *Curator: The Museum Journal* (online first). https://doi.org/10.1111/cura.70012

Pradel, F. (2021). Biased representation of politicians in Google and Wikipedia search? The joint effect of party identity, gender identity and elections. *Political Communication*, *38*(4), 447-478.

Price, B. (2018). Material memory: The politics of nostalgia on the eve of MAGA. *American Studies, 57*(1), 103-115.

Otterbacher, J., Bates, J. & Clough, P. (2017). Competent men and warm women: Gender stereotypes and backlash in image search results. In *Proceedings of the 2017 CHI Conference on Human Factors in Computing Systems* (pp. 6620-6631). ACM.

Reid, E. (2025, May 20). AI in Search: Going beyond information to intelligence. https://blog.google/products/search/google-search-ai-mode-update/#ai-mode-search

Rohrbach, T., Makhortykh, M., & Sydorova, M. (2024). *Finding the white male: The prevalence and consequences of algorithmic gender and race bias in political Google searches*. arXiv. https://doi.org/10.48550/arXiv.2405.00335

Rohrbach, T., Makhortykh, M., & Sydorova, M. (2025). *Campaigning through the lens of Google: A large-scale algorithm audit of Google searches in the run-up to the Swiss Federal Elections 2023*. arXiv. https://doi.org/10.48550/arXiv.2507.06018

Ross, L., & Tajima, A. (2024). Yellow star re appropriation: Revitalized Nazi symbolism in anti-Covid vaccine protests on social media. *Journal of Contemporary Antisemitism, 7*(1), 53-70.

Scheuerman, M., Paul, J., & Brubaker, J. R. (2019). How computers see gender: An evaluation of gender classification in commercial facial analysis services. *Proceedings of the ACM on Human-Computer Interaction, 3*, 1-33.

Schultheiß, S., & Lewandowski, D. (2023). Misplaced trust? The relationship between trust, ability to identify commercially influenced results and search engine preference. *Journal of Information Science, 49*(3), 609-623.

Sodaro, A. (2019). Selective memory: Memorial museums, human rights, and the politics of victimhood. In J. Apsel & Amy Sodaro (Eds.), *Museums and sites of persuasion: Politics, memory and human rights* (pp. 19-35). Routledge.

Sontag, S. (2003). *Regarding the pain of others*. Picador.

Statcounter. (n.d.). Search engine market share worldwide. Retrieved November 25, 2025, from https://gs.statcounter.com/search-engine-market-share

Steinweis, A. E. (2005). The Auschwitz analogy: Holocaust memory and American debates over intervention in Bosnia and Kosovo in the 1990s. *Holocaust and Genocide Studies*, 19(2), 276-289.

Stoet, G., & Geary, D. C. (2018). The gender-equality paradox in science, technology, engineering, and mathematics education. *Psychological Science*, 29(4), 581-593.

Ulloa, R., Richter, A., Makhortykh, M., Urman, A. & Kacperski, C. (2024). Representativeness and face-ism: Gender bias in image search. *New Media & Society*, 26(6), 3541-3567.

Urman, A., Makhortykh, M., & Ulloa, R. (2022). Auditing the representation of migrants in image web search results. *Humanities and Social Sciences Communications*, 9(1), 1-16.

Urman, A., & Makhortykh, M. (2023). You are how (and where) you search? Comparative analysis of web search behavior using web tracking data. *Journal of Computational Social Science, 6*(2), 741-756.

Urman, A., & Makhortykh, M. (2024). "Foreign beauties want to meet you": The sexualization of women in Google's organic and sponsored text search results. *New Media & Society*, *26*(5), 2932-2953.

Urman, A., Makhortykh, M., & Hannak, A. (2025). WEIRD audits? Research trends, linguistic and geographical disparities in the algorithm audits of online platforms - a systematic literature review. In *Proceedings of the 2025 ACM Conference on Fairness, Accountability, and Transparency* (pp. 375-390). ACM.

USHMM. (2025). Jewish losses during the Holocaust: By country. Retrieved November 25, 2025, from https://encyclopedia.ushmm.org/content/en/article/jewish-losses-during-the-holocaust-by-country

Vargas-Veleda, Y., del Mar Rodríguez-González, M., & Marauri-Castillo, I. (2025). Visual representations in AI: A study on the most discriminatory algorithmic biases in image generation. *Journalism and Media, 6*(3), 1-15.

Vianna, B. C. (2025). Aesthetic biases and opacity tactics in the training of visual artificial intelligence models. In *International Conference on Computational Intelligence in Music, Sound, Art and Design* (pp. 278-293). Springer.

Vlasceanu, M., & Amodio, D. M. (2022). Propagation of societal gender inequality by internet search algorithms. *Proceedings of the National Academy of Sciences*, *119*(29), 1-8.

WEC. (2024, June 11). *Global gender gap report 2024*. https://www.weforum.org/publications/global-gender-gap-report-2024/digest/

Weimann, G., & Masri, N. (2023). Spreading Hate on TikTok. *Studies in Conflict & Terrorism*, 46(5), 752–765.

Zelizer, B. (2001). Gender and atrocity: Women in Holocaust photographs. In B. Zelizer (Ed.), *Visual culture and the Holocaust* (pp. 247-275). Rutgers University Press.

Zhu, Y. et al. (2025). Large language models for information retrieval: A survey. *ACM Transactions on Information Systems*, 44(1), 1-54.

**Online supplement**

**Table 1**. The list of locations and queries for the audit, together with the number of images used for the analysis for each location-query pair

| Country | Language | Query | Image count |
|---|---|---|---|
| Argentina | Spanish | holocausto victimas | 50 |
| Argentina | Spanish | holocausto víctima | 50 |
| Australia | English | holocaust victims | 50 |
| Australia | English | holocaust victim | 50 |
| Austria | German | holocaust opfer | 50 |
| Belarus | Belarusian | халакост ахвяра | 50 |
| Belarus | Russian | холокост жертвы | 50 |
| Belarus | Russian | холокост жертва | 50 |
| Belarus | Belarusian | халакост ахвяры | 50 |
| Belgium | Dutch | holocaust slachtoffers | 50 |
| Belgium | French | holocauste victime | 50 |
| Belgium | German | holocaust opfer | 50 |
| Belgium | French | holocauste victimes | 50 |
| Belgium | Dutch | holocaust slachtoffer | 50 |
| Brazil | Portuguese | holocausto vítima | 50 |
| Brazil | Portuguese | holocausto vítimas | 50 |
| Bulgaria | Bulgarian | холокост жертва | 50 |
| Bulgaria | Bulgarian | холокост жертви | 50 |
| Canada | French | holocauste victimes | 50 |
| Canada | French | holocauste victime | 50 |
| Canada | English | holocaust victims | 50 |
| Canada | English | holocaust victim | 50 |
| Costa Rica | Spanish | holocausto victimas | 50 |
| Croatia | Croatian | holokaust žrtva | 50 |
| Croatia | Croatian | holokaust žrtve | 50 |
| Czech Republic | Czech | holocaust oběť | 50 |
| Czech Republic | Czech | holocaust obětí | 50 |
| Denmark | Danish | holocaust offer | 50 |
| Denmark | Danish | holocaust ofre | 50 |
| France | Arabic | ضحية المحرقة | 50 |
| France | French | holocauste victime | 50 |
| France | French | holocauste victimes | 50 |
| France | Arabic | ضحايا المحرقة | 50 |

| Germany | Arabic | ضحايا المحرقة | 50 |
|---|---|---|---|
| Germany | German | holocaust opfer | 50 |
| Germany | Arabic | ضحية المحرقة | 50 |
| Greece | Greek | ολοκαύτωμα θύμα | 50 |
| Greece | Greek | ολοκαύτωμα θύματα | 50 |
| Hungary | Hungarian | holokauszt áldozatok | 50 |
| Hungary | Hungarian | holokauszt áldozat | 50 |
| Israel | Arabic | ضحايا المحرقة | 50 |
| Israel | Arabic | ضحية المحرقة | 50 |
| Israel | Hebrew | קורבן השואה | 50 |
| Israel | Hebrew | קורבנות השואה | 50 |
| Italy | Italian | olocausto vittima | 50 |
| Italy | Italian | olocausto vittime | 50 |
| Japan | Japanese | ホロコーストの犠牲者 | 50 |
| Latvia | Latvian | holokausts upuri | 50 |
| Latvia | Latvian | holokausts upuris | 50 |
| Netherlands | Dutch | holocaust slachtoffers | 50 |
| Netherlands | Dutch | holocaust slachtoffer | 50 |
| Norway | Norwegian | holocaust ofre | 50 |
| Norway | Norwegian | holocaust offer | 50 |
| Peru | Spanish | holocausto víctima | 50 |
| Peru | Spanish | holocausto victimas | 50 |
| Poland | Polish | zagłada żydów ofiara | 50 |
| Poland | Polish | zagłada żydów ofiary | 50 |
| Poland | Ukrainian | голокост жертва | 31 |
| Poland | Ukrainian | голокост жертви | 50 |
| Romania | Romanian | holocaust victime | 50 |
| Romania | Romanian | holocaust victimă | 50 |
| Russia | Russian | холокост жертвы | 50 |
| Russia | Russian | холокост жертва | 50 |
| Saudi Arabia | Arabic | ضحية المحرقة | 50 |
| Saudi Arabia | Arabic | ضحايا المحرقة | 50 |
| Serbia | Serbian | холокоста жртва | 50 |
| Serbia | Serbian | холокоста жртве | 50 |
| Slovenia | Slovenian | holokavst žrtev | 50 |
| Slovenia | Slovenian | holokavst žrtve | 50 |
| Sweden | Swedish | förintelsens offer.htm | 50 |
| Sweden | Swedish | förintelsens offer | 50 |
| Switzerland | French | holocauste victime | 50 |

| Switzerland | Italian | olocausto vittima | 50 |
|---|---|---|---|
| Switzerland | Italian | olocausto vittime | 50 |
| Switzerland | German | holocaust opfer | 50 |
| Switzerland | French | holocauste victimes | 50 |
| Turkey | Turkish | holokost kurban | 50 |
| Turkey | Turkish | holokost kurbanlar | 50 |
| UK | English | holocaust victims | 50 |
| UK | English | holocaust victim | 50 |
| Ukraine | Ukrainian | голокост жертва | 50 |
| Ukraine | Russian | холокост жертва | 50 |
| Ukraine | Russian | холокост жертвы | 50 |
| Ukraine | Ukrainian | голокост жертви | 50 |
| US | English | holocaust victims | 50 |
| US | Spanish | holocausto víctima | 50 |
| US | English | holocaust victim | 50 |
| US | Spanish | holocausto victimas | 50 |